\documentclass[prl,showpacs,twocolumn,amsmath,amssymb,superscriptaddress]{revtex4-1}

\usepackage{color}
\usepackage{graphicx}

\definecolor{kg}{rgb}{0,0,1} 
\definecolor{cm}{rgb}{.6,0.1,.6} 

\newcommand{\ket}[1]{|#1\rangle}

\begin{document}

\title[Hybridization between carbon nanotubes]
{Spin-dependent electronic hybridization in a rope of carbon nanotubes}

\author{Karin Go\ss}
\affiliation{{Peter Gr\"unberg Institut}, Forschungszentrum J\"ulich \& JARA J\"ulich Aachen Research Alliance, 52425 J\"ulich, Germany}
\author{Sebastian Smerat}
\affiliation{Physics Department, Arnold Sommerfeld Center for Theoretical Physics, Ludwig-Maximilians-Universit\"at M\"unchen, 80333 M\"unchen, Germany}
\author{Martin Leijnse}
\affiliation{Niels Bohr Institute \& Nano-Science Center,
University of Copenhagen, Universitetsparken 5, 2100 Copenhagen,
Denmark}
\author{Maarten R. Wegewijs}
\affiliation{{Peter Gr\"unberg Institut}, Forschungszentrum J\"ulich \& JARA J\"ulich Aachen Research Alliance, 52425 J\"ulich, Germany}
\affiliation{{Institute for Theory of Statistical Physics}, RWTH Aachen, 52056 Aachen, Germany}
\author{Claus M. Schneider}
\affiliation{{Peter Gr\"unberg Institut}, Forschungszentrum J\"ulich \& JARA J\"ulich Aachen Research Alliance, 52425 J\"ulich, Germany}
\author{Carola Meyer}
\affiliation{{Peter Gr\"unberg Institut}, Forschungszentrum J\"ulich \& JARA J\"ulich Aachen Research Alliance, 52425 J\"ulich, Germany}
\email{c.meyer@fz-juelich.de}

\date{\today}

\begin{abstract}
We demonstrate single electron addition to different strands of a carbon nanotube rope. Anticrossings of anomalous conductance peaks occur
in quantum transport measurements through the parallel quantum dots forming on the individual strands. We determine the magnitude and the sign of the hybridization as well as the Coulomb interaction between the carbon nanotube quantum dots, finding that the bonding states dominate the transport.
In a magnetic field the hybridization is shown to be selectively suppressed due to spin effects.
\end{abstract}
\maketitle
Molecular electronics and spintronics aim at exploiting the chemical versatility of molecules to control charge and magnetism  in nanoscale devices. However, the assembly of molecular structures in junctions for electric and magnetic manipulation is a challenging task \cite{Bogani2010,Osorio2008,Ghosh2005,Park1999}. Carbon nanotubes (CNTs) are particularly promising as building blocks of new devices for nano-electronics \cite{Steele2009-2,Lassagne2009,Huttel2009-2,Stampfer2006}, which exhibit interesting spin properties \cite{Kuemmeth2008,Jarillo2005,Tombros2006,Schoenenberger2005} and can be useful for quantum information processing \cite{Meyer2007,Churchill2009}. Fundamental aspects of single-molecule devices require an understanding of strong perturbations by environmental effects, e.\,g., the interaction with contacts or neighboring molecules \cite{Osorio2010}. These interactions can, in principle, be studied on a single molecule level using scanning probe techniques as for instance scanning near-field optical microscopy \cite{Betzig1993}, tip-enhanced Raman spectroscopy \cite{Hayazawa2001,Hartschuh2003} or scanning tunneling spectroscopy (STS) \cite{Ouyang2001}. However, in-situ characterization of actual devices{, e.\,g., field-effect transistors,} is difficult to implement and only STS can detect spin dependent phenomena.

As an alternative approach, one may exploit the differential electrostatic gating effect, which was found to occur for contacted CNTs filled with fullerenes \cite{Eliasen2010} and for single molecules in nanojunctions \cite{Osorio2010}. In this respect, bundled CNTs are interesting: within a rope one expects the strands to be at a different potential and to respond differently to the external electric fields due to electrostatic effects \cite{Kaasbjerg2008}. Low-temperature transport spectroscopy is sensitive to these potential variations on the sub-meV scale, allowing the study of interactions between coupled nanoscale conductors.

In this Letter, we show that transport spectroscopy can resolve both charge addition to individual strands of a single CNT rope {as well as the coupling between these strands caused by molecular interactions}. We determine the hybridization and the electrostatic interaction between {parallel} quantum dots (QDs) forming on the different strands. We extract both the magnitude and sign of the hybridization and find that current transport occurs via the bonding states of the coupled QD system. Furthermore, {by applying a magnetic field} the electronic hybridization is selectively suppressed due to spin effects. This offers prospects for accessing individual charge and spin degrees of freedom in coupled carbon-based molecular systems.

\begin{figure}[htb]
  \includegraphics{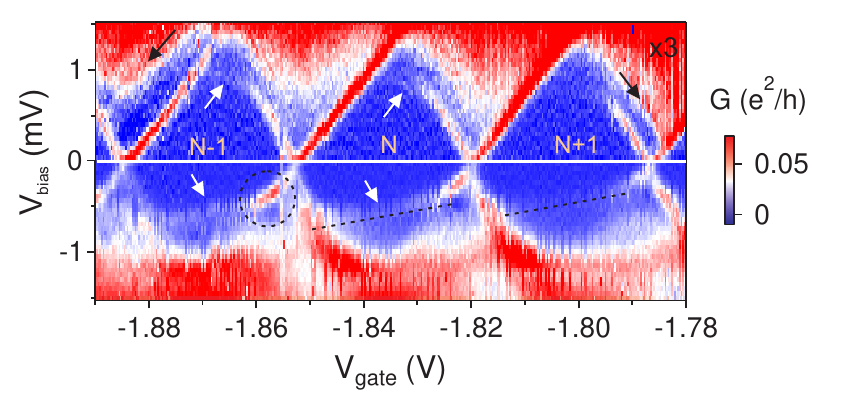}
  \caption{(color online) The differential conductance plotted versus gate and bias voltage showing {three} complete Coulomb diamonds. {Black arrows indicate excited-state resonances.} Secondary resonances, indicated by white arrows and dashed lines, cross the region of Coulomb blockade. A circle indicates the anticrossing. To clearly display all features, the conductance at positive bias {voltage} has been multiplied by three.}
  \label{fig:diamonds}
\end{figure}

The CNTs of the reported device were grown on a $\mathrm{Si/SiO_2}$ substrate by chemical vapor deposition at 920$^\circ$C using a Fe/Mo catalyst and methane as precursor \cite{Kong1998}. At these temperatures the process results mainly in single-walled CNTs and few double-walled CNTs \cite{Spudat2009}. Source and drain electrodes (5\,nm Ti/60\,nm Au) {were} patterned by electron beam lithography to form a quantum dot of length 360\,nm with the highly doped Si substrate acting as back gate. The height profile of an atomic force micrograph of the QD region shows a CNT height of $\sim$7\,nm evidencing that the device consists of a CNT rope rather than an individual tube.

At room temperature the device shows metallic behavior with a resistance of 290\,k$\Omega$. Low-temperature transport properties were measured in a dilution refrigerator at a base temperature of $\sim$30\,mK. In the plot of the differential conductance (Fig.~\ref{fig:diamonds}) Coulomb blockade in a QD is observed by the typical diamond shaped signatures {together with excited states, which imply size quantization.}

The diamonds close at zero bias voltage and {are observed for a large range of the gate voltage}
(-2.4\,V to +1.3\,V), showing that a stable QD is formed in the rope. {Although the diamonds vary in size, no regular pattern of shell-filling, like the fourfold pattern typical for individual single-walled CNTs \cite{Sapmaz2005}, is found. }
{More importantly, }two salient features are observed in Fig.~\ref{fig:diamonds}. First, additional conductance peaks appear within the region of Coulomb blockade. These secondary resonances exhibit a weak gate voltage dependence (small slope) and do not appear symmetrically at positive and negative bias voltage. {Proceeding from one Coulomb diamond to the next, the positions of these resonances jump in voltage (dashed lines).}

Second, anticrossings are observed whenever the secondary resonances meet with a main resonance of the Coulomb diamonds {with} the same slope. In the vicinity of these points the secondary resonances show an enhanced conductance. Inelastic co-tunneling \cite{DeFranceschi2001} cannot explain the combination of these features {considering} excitations of {only} a single QD.

\begin{figure}[htb]
  \includegraphics{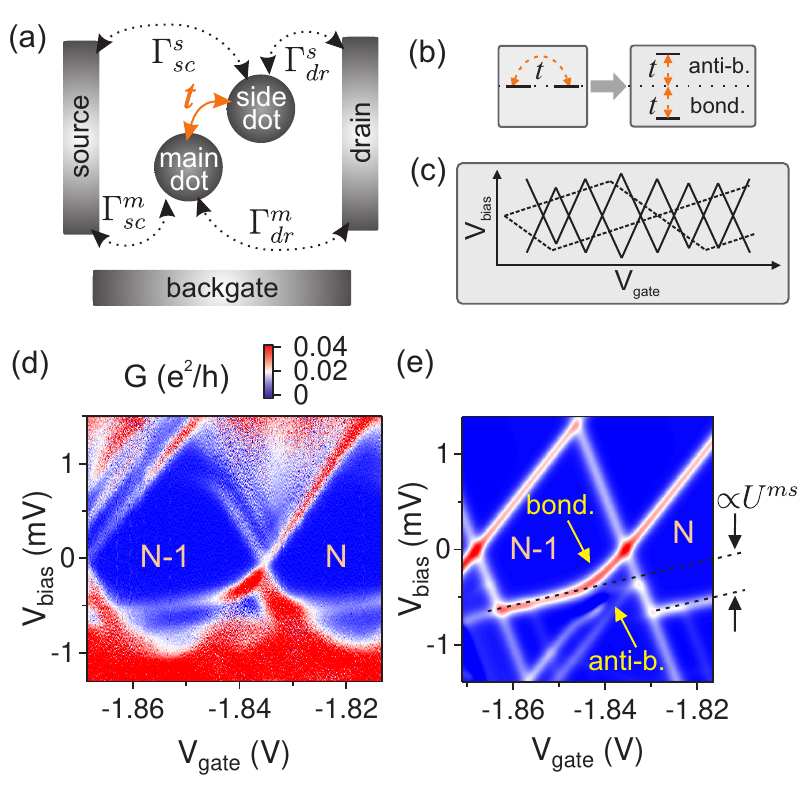}
  \caption{(color online) \emph{(a)}\,Sketch of the transport model: Two QDs are contacted in parallel with different gate coupling strength. Electrons tunnel with rates $\Gamma$ from the leads into and out of the QDs. The QDs are connected due to hybridization with an amplitude $t$. \emph{(b)}\,The hybridization of two orbitals generates a bonding and an anti-bonding eigenstate. \emph{(c)}\,{Sketch of two overlaying diamond patterns formed by two differently gate coupled, parallel QDs. \emph{(d)}\,High resolution measurement of the $(N-1)$- and $N$-diamond of Fig.~\ref{fig:diamonds}.} \emph{(e)}\,Calculated stability diagram {using the model of \emph{(a)}}.}
  \label{fig:model}
\end{figure}

Instead, the data indicate that several coupled QDs are formed in the rope and contacted in parallel. In order to explain the findings above, we use the model \cite{Eliasen2010} sketched in Fig.~\ref{fig:model}a. {In this model,} two QDs are contacted in parallel with a different gate coupling strength for the different dots. This leads to a $dI/dV_{bias}$-diagram with Coulomb diamonds, which originate from a QD referred to as the main dot (indexed $m$) from here on. {In addition, resonances with a smaller slope are seen as part of a second diamond pattern, which overlays the pattern of the main dot (see Fig.~\ref{fig:model}c). These indicate a weakly gate-coupled side dot (indexed $s$) formed in a different CNT strand. Due to this differential gating effect, charges can be added selectively to the parallel QDs in the CNT rope. In contrast to serial double QDs \cite{Mason2004,Sapmaz2006,Graeber2006}, transport is possible even if one of the dots is in Coulomb blockade. This enables a detailed spectroscopy of the hybridization between strands, which practically can be addressed only with a single tunable gate.}

Standard master equations are used for transport calculations, accounting for {the} lowest order tunnel processes to the leads.
The parallel double QD is described within a constant interaction model \cite{Oreg2000}, extended to account for a finite hybridization integral between many-body states of the two QDs. The electrochemical potential $\mu_{\nu}^i$ for adding an electron to orbital $\nu$ on dot $i = m,s$ depends on the initial many-body state of the system (i.\,e., before adding the electron) but always satisfies the proportionality relation:
\begin{equation}
	\mu_{\nu}^i \propto -|e| \alpha_{sc}^i V_{bias} - |e| \alpha_{gt}^i V_{gate},
\end{equation}
where $\alpha^i_{sc,dr,gt} = {{C_{sc,dr,gt}^i}/{C^i}}$ is the capacitive coupling strength of dot $i$ to the source, drain and gate electrode and $C^i = C_{sc}^i + C_{dr}^i + C_{gt}^i$ is the capacitance of the system.

The model also accounts for a capacitive coupling between the two QDs through the inter-dot charging energy $U^{ms}${, which can be resolved using the differential gating}. With each Coulomb diamond, proceeding in the positive $V_{gate}$ direction, the main dot is charged with an additional electron. This leads to a {discrete change} in the electrostatic potential on the side dot and thus to an energy offset in the stability diagram. If the hybridization were negligible compared to thermal and tunnel broadening, the conduction lines should show a crossing at the diamond edges, as observed for parallel contacted molecules by Osorio et al. \cite{Osorio2010}. On the other hand, if the hybridization amplitude $t$ of the two orbitals is significant on the scale of the energy difference between the two orbitals, anticrossings of excitation lines should appear.
At resonance, hybridized bonding $\ket{-}$ and anti-bonding $\ket{+}$ states are generated, which are split in energy by $2|t|$ (see Fig.~\ref{fig:model}b). Figure~\ref{fig:model}e shows that the calculations reproduce the experimentally observed features {of the high resolution measurement in Fig.~\ref{fig:model}d very well, i.e. secondary resonances, which cross the Coulomb blocked regions and exhibit an anticrossing as well as an energy offset. The calculations exhibit only thermal broadening, while tunnel broadening is neglected. Also they include only one orbital for each dot. In the experiment, the presence of additional orbitals wash out the features at high negative bias voltage.}
{In Fig.~\ref{fig:model}e, the bonding and anti-bonding states clearly contribute} very differently to the conductance. This relates to the sign of $t$, as shown in the following.

\begin{figure}[htb]
  \includegraphics{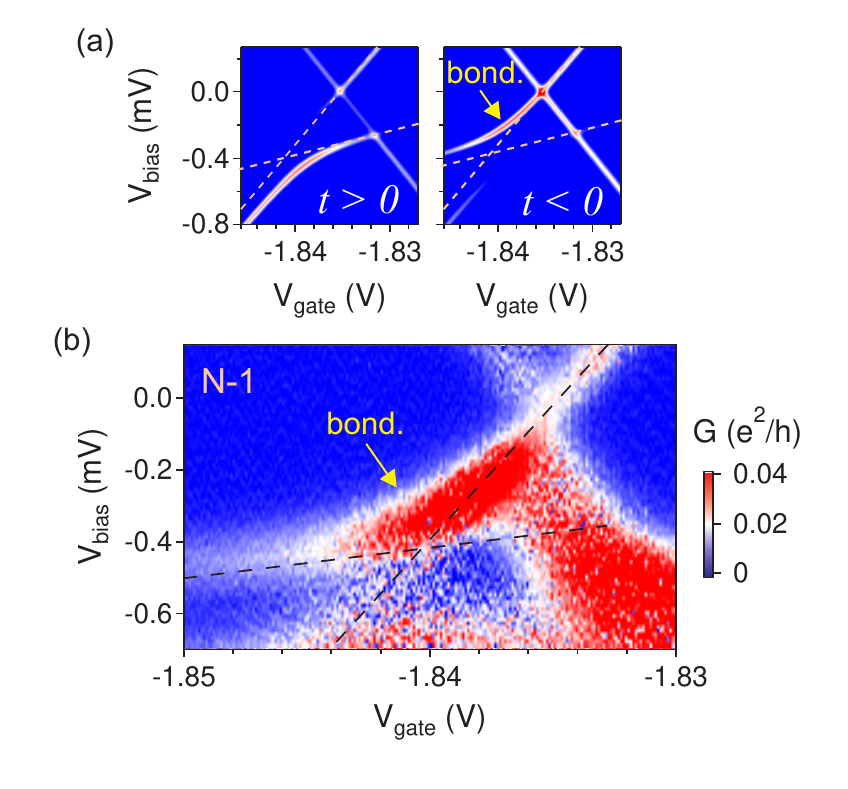}
  \caption{(color online) \emph{(a)}\,Calculated stability diagrams showing anticrossings for opposite signs of the hybridization integral $t$. {The dashed lines indicate the resonance positions for the unhybridized states.} \emph{(b)}\,{Close-up of the anticrossing at negative bias voltage in Fig.~\ref{fig:model}d.}}
  \label{fig:hybridization}
\end{figure}

The hybridized eigenstates {for a single electron in the coupled QD system} are
 \begin{equation}
 \begin{split}
 	\ket{+}  = & \phantom{-} \cos{\theta} \ket{\text{m}} + \sin{\theta} \ket{\text{s}}\\
 	\ket{-}   =  & -\sin{\theta} \ket{\text{m}} + \cos{\theta} \ket{\text{s}},
 \end{split}
 \end{equation}
where $\ket{\text{m}}$ and $\ket{\text{s}}$ are the original states on the main dot and the side dot with the energies $\pm \epsilon = \pm \epsilon(V_{bias}, V_{gate})$ taken relative to their resonance energy. The hybridized states depend on the applied voltages through $\theta$, where
$\tan \theta = \frac{\sqrt{\epsilon^2 + t^2} - \epsilon}{t}$, and their corresponding eigenenergies are $\pm \sqrt{\epsilon^2 + t^2}$. This analysis can be extended to more electrons in the system. The rate for the tunneling of an electron from the leads into the $\ket{\pm}$ state is proportional to the corresponding tunnel matrix elements $T_{\pm}$:
\begin{equation*}
    \Gamma_{\pm}  \propto
        |T_{\pm}|^2 = T_{\text{m/s}}^2 \cos^2{\theta} + T_{\text{s/m}}^2 \sin^2{\theta} \pm T_{\text{m}} T_{\text{s}} \cos{\theta} \sin{\theta}, \quad \\
\end{equation*}
Here $T_{\text{m}}$ and $T_{\text{s}}$ are the matrix elements for tunneling into $\ket{\text{m}}$ or $\ket{\text{s}}$, respectively, whose sign we take to be the same since the strands form a single junction.

The sign of the hybridization $t$ determines which tunneling rate {(into the bonding or into the anti-bonding state)} will be suppressed or enhanced.
In Fig.~\ref{fig:hybridization}a the left (right) panel shows a calculated anticrossing of two hybridizing states with a positive (negative) hybridization integral $t$, where a clearly enhanced anti-bonding (bonding) state is visible. {In both cases, the conductance in the enhanced state exhibits an additional pronounced gate voltage dependence, which is characteristic for the hybridization.}
{The interference effect at the anticrossing is due to an interference of two bound states, both connected to the leads (see Fig.~\ref{fig:model}a,b), and is hence distinct from the Fano effect, where one bound state interferes with an unscattered wave. We find an enhanced current of the bonding state at the anticrossing in Fig.~\ref{fig:hybridization}b.} Thus, the CNT strands hybridize with a negative hybridization integral, i.e. the wave functions of the QD states overlap with the same sign.

{Figure~\ref{fig:model}d exhibits additional secondary resonances at positive bias voltages, which are less visible at zero magnetic field.} From the magnitude of the gap at the anticrossings we can estimate {$t_{neg}  \approx -0.075$\,meV at negative bias voltage and $t_{pos} \approx -0.1$\,meV} at positive bias voltage. We obtain the inter-dot charging energy from the energy offset of the secondary resonances as $U^{ms}_{neg} \approx 0.2$\,meV { and $U^{ms}_{pos} \approx 0.4$\,meV.} From the different values for the hybridization integral and the inter-dot charging energy at {negative and positive} bias voltage, we deduce that these resonances belong to \emph{two} additional side dots which interact differently with the main dot. These side dots are formed in different CNT strands of the rope.

{Modeling each anticrossing with a single side dot is valid, since we find no indication in the measurements for an interaction between the side dots and the data are reproduced very well.} Due to a large source (drain) coupling only the negative (positive) slope of the secondary resonance for the two different dots can be observed. {The magnitude of the anticrossing gap and the energy offset both depend on the bias coupling. By comparison with several model calculations we find $\alpha^{s}_{sc,dr} \approx 0.7 - 1.0$. Our estimates for $t$ and $U^{ms}$ are thus upper bounds but with the right order of magnitude.} Moreover, they compare favorably with the hybridization strength and inter-dot charging energy found for fullerenes hybridizing with a CNT in a peapod~\cite{Eliasen2010}.

\begin{figure}[htb]
  \includegraphics{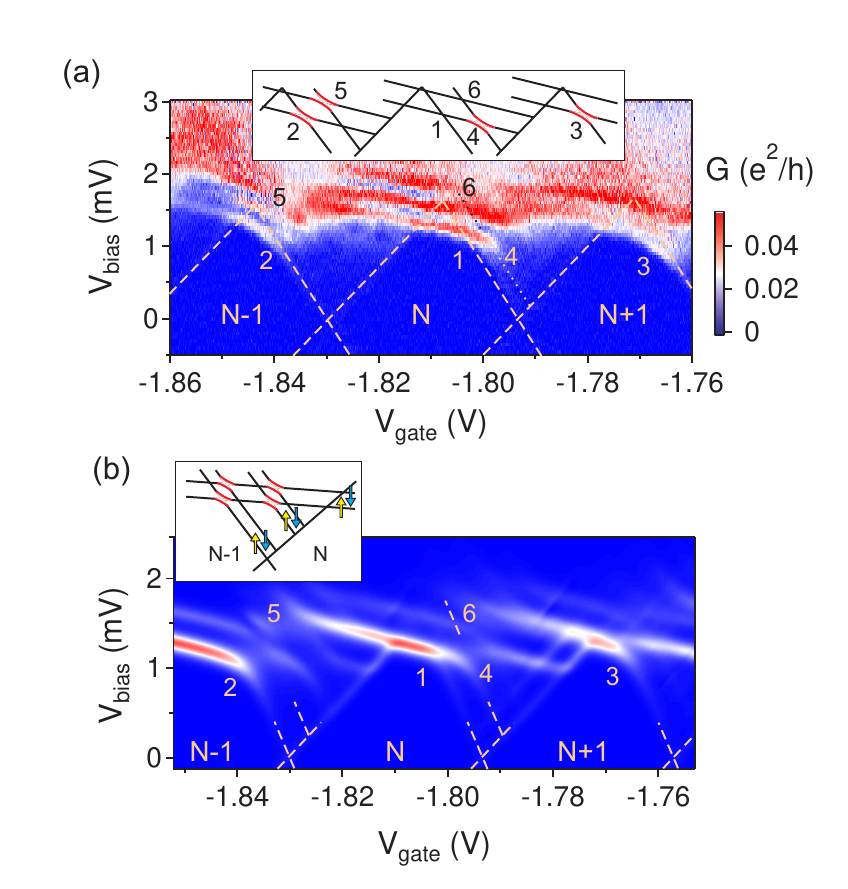}
  \caption{(color online) \emph{(a)}\,Crossings and anticrossings at $B=$\,10\,T (same charge states of the main dot as in zero field). Dashed lines indicate the extrapolated diamond edges. \emph{Inset:}\,Guide to the measurements. The anti-bonding states are included for completeness. \emph{(b)}\,Calculated stability diagram with $N$\,$=$\,1 according to the model sketched in the inset. Coupling parameters are taken from the experiment and tunneling rates are adjusted accordingly. \emph{Inset:}\,Schematic stability diagram for different states on the QDs. Arrows indicate spin split states.}
  \label{fig:B-zoom-model}
\end{figure}

In order to investigate a possible influence of the spin degrees of freedom on the hybridization within the rope, the transport spectrum was measured in an applied magnetic field. {Compared to the measurements at $B=$~0\,T, the transport through the main dot is strongly suppressed at $B=$~10\,T (Fig.~\ref{fig:B-zoom-model}a) and} the secondary resonances at positive bias voltage appear as the most prominent feature.

Again, the anticrossings show an enhancement of the bonding state. The energy offset, hybridization integral, and gate coupling are the same as in zero field, evidencing that these resonances indeed originate from the same CNT strand as the ones at $B=$ 0\,T.

Furthermore, Fig.~\ref{fig:B-zoom-model}a shows additional resonances involving  excited states of the side dot, which exhibit the same weak gate voltage dependence and anticrossings as the initial secondary resonances. Amidst these anticrossings a clear \emph{crossing} (marked (1)) appears , when the chemical potential $\mu^s$ for adding an electron to the ground state of the side dot is at resonance with the chemical potential of the main dot $\mu^{m}$ for the ground state transition from $N$ to $N+1$ electrons.
This crossing reveals that the resonant states of the coupled QD system containing in total $N+1$ electrons have different quantum numbers, most probably different spins, preventing the states from hybridizing.
In Fig.~\ref{fig:B-zoom-model}b we present a model calculation which assumes a  state on the side dot and an additional orbital state on the main dot. All of the states are spin split. {The tunneling rates are adjusted according to the observation of an overall suppressed transport through the main dot.}

The selective suppression of the hybridization at position (1) and the ground state transitions marked (2) and (3) which do hybridize are reproduced very well.
Similarly, both the calculation and the measurement show that the transition involving excited states at position (6) also exhibits a crossing,
in contrast to the  anticrossing  excitations at positions (4) and (5).
This is consistent with the crossing for the ground-state transition discussed above. Using gate and bias voltages, it is thus possible to select spin states, which do not hybridize.

{In summary, we measured quantum transport through several parallel quantum dots, formed in different carbon nanotube strands within a rope. Using differential gating, we determined the magnitude and the sign of the electronic hybridization between the states of the coupled quantum dots and found that the transport is enhanced when electrons tunnel via the bonding states. In a magnetic field, the hybridization between these many-body states is found to be selectively suppressed by spin effects. We have thus shown that the molecular hybridization within a CNT rope can be detected and manipulated both by electric and magnetic fields. Such tunability of the hybridization is a key element in accessing localized charges and spins in coupled molecular systems, realized also in, e.g., graphene or single molecules.}

\begin{acknowledgments}
We thank H.\ Schoeller and U. Schollw\"ock for fruitful discussions and S. Trellenkamp for e-beam writing, as well as S.\ Est\'{e}vez Hern\'{a}ndez, H.\ Kertz, H.\ Pfeifer, W.\ Harneit and J.\ Lauer for technical assistance. We acknowledge the DFG (FOR 912) and the EU under the FP7 STREP program SINGLE for funding.
\end{acknowledgments}

\end{document}